\documentclass[reprint,amssymb, amsmath, aps, superscriptaddress, showpacs, footinbib, prb]{revtex4-2}
\usepackage{graphicx,epstopdf} 
\usepackage{amsmath}
\usepackage{xcolor, soul}

\usepackage[colorlinks,bookmarks=false,citecolor=black,linkcolor=black,urlcolor=black]{hyperref}
\usepackage{amssymb}
\usepackage{longtable}
\usepackage{soul}

\newcommand{\be}{\begin{equation}}
\newcommand{\ee}{\end{equation}}
\newcommand{\bea}{\begin{eqnarray}}
\newcommand{\eea}{\end{eqnarray}}
\newcommand{\bse}{\begin{subequations}}
\newcommand{\ese}{\end{subequations}}

\begin{document}

\title{Intra-unitcell cluster-cluster magnetic compensation and large exchange bias in cubic alloys}

\author{Bimalesh Giri}
\thanks{authors contributed equally}
\affiliation{School of Physical Sciences, National Institute of Science Education and Research, HBNI, Jatni-752050, India}
\author{Bhawna Sahni}
\thanks{authors contributed equally}
\affiliation{Department of Physics, Indian Institute of Technology Bombay, Mumbai- 400076, India}
\author{C.\ Salazar Mej\'{i}a}
\thanks{authors contributed equally}
\affiliation{Dresden High Magnetic Field Laboratory (HLD-EMFL), Helmholtz-Zentrum Dresden-Rossendorf, 01328 Dresden, Germany}
\author{S. Chattopadhyay}
\affiliation{Dresden High Magnetic Field Laboratory (HLD-EMFL), Helmholtz-Zentrum Dresden-Rossendorf, 01328 Dresden, Germany}
\author{Uli Zeitler}
\affiliation{High Field Magnet Laboratory (HFML-EMFL), Radboud University,Toernooiveld 7, 6525 ED Nijmegen, The Netherlands}
\author{Aftab Alam}
\email{aftab@phy.iitb.ac.in}
\affiliation{Department of Physics, Indian Institute of Technology Bombay, Mumbai- 400076, India}
\author{Ajaya K. Nayak}
\email{ajaya@niser.ac.in}
\affiliation{School of Physical Sciences, National Institute of Science Education and Research, HBNI, Jatni-752050, India}

\begin{abstract}

Composite quantum materials  are the ideal examples of multifunctional systems which simultaneously host more than one novel quantum phenomenon in physics. Here, we present a combined theoretical and experimental study to demonstrate the presence of an extremely large exchange bias in the range 0.8~T-2.7~T and a fully compensated magnetic state (FCF) in a special type of Pt and Ni doped Mn$ _{3} $In cubic alloy. Here, oppositely aligned uncompensated moments in two different  atomic clusters sum up to zero which are responsible for the FCF state.  Our dDensity functional theory (DFT) calculations show the existence of several possible ferrimagnetic configurations with the FCF as the energetically most stable one. The microscopic origin of the large exchange bias can be interpreted in terms of the exchange interaction between the FCF background and the uncompensated ferrimagnetic clusters stabilized due to its negligible energy difference with respect to the FCF phase. We utilize pulsed  magnetic field up to 60 T and 30~T static field magnetization measurements to confirm the intrinsic nature of exchange bias in our system. Finally, our Hall effect measurements demonstrate the importance of uncompensated noncoplanar interfacial moments for the realization of large EB. The present finding of gigantic exchange bias in a unique compensated ferrimagnetic system opens up a direction for the design of novel quantum phenomena for the technological applications.

\end{abstract}

\maketitle

\section{\textbf{Introduction}}

Coexistence of two or more complimentary quantum phenomena in a single material often provides a fertile ground to explore the fundamental correlation between these different events in physics. Apart from basic science, such materials displaying conjugation of different quantum properties, called the composite quantum materials, can also open the door for potential technological applications \cite{Tokura17}. Large exchange bias (EB) and fully compensated ferrimagnets (FCF) are two distinct quantum phenomena which can be connected via the  common prerequisite of a special type of exchange interaction. EB, which is represented by an asymmetrical offset in the magnetic hysteresis loop, is a measure of unidirectional exchange anisotropy in an exchange coupled magnetic system \cite{1,Nogues1999,Giri2011,Wang2011,Nayak2013}. EB effect has been studied extensively due to its utmost importance in the field of spintronics, e.g. setting up pinning layer in  the giant magnetoresistance (GMR) based devices \cite{1991,Gider1998,Binek2005,Furubayashi2008}, beating the super-paramagnetic limit  in magnetic nano-particle based ultra–high density recording media \cite{9} etc. The EB phenomenon has been studied ubiquitously in the ferromagnetic(FM)-ferrimagnet(FiM) \cite{6}, FM-spin glass (SG) \cite{7}, antiferromagnetic (AFM)-FiM \cite{8}, and AFM-SG \cite{Maniv2021} systems. Although various models have been proposed to understand the origin of EB \cite{10,Takano1997,Milteyni2000,Nowak2002}, an essential requirement of most of the microscopic models is the presence of uncompensated AFM moment  at the interface. Hence, the AFM layer plays a crucial role to induce the interface exchange field (H$ _{E}$) and causes the asymmetry in the hysteresis loop. The  H$ _{E} $, which is in general inversely proportional to the magnetization(M$_{FM}$) of the ferromagnet, can be controlled by tuning M$_{FM}$ and the nature of interface. 
 \begin{figure*}[tb!]
 	\begin{center}
 		\includegraphics[angle=0,width=17 cm,clip=true]{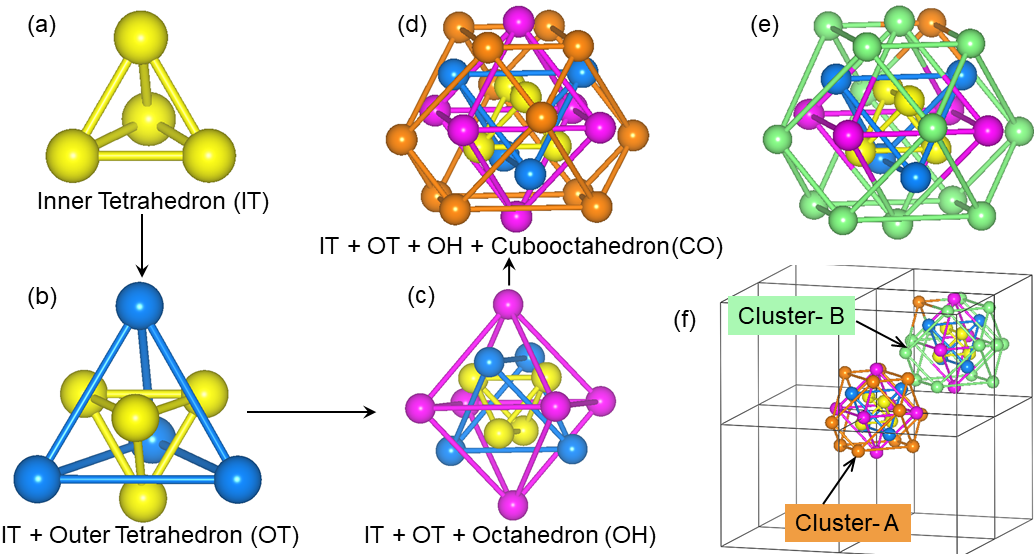}
 		\caption{\label{FIG1} Decomposition of the Mn$ _{3} $In crystal structure. (a) Four Mn atoms (yellow spheres) forming an inner tetrahedron (IT). (b) Four Mn atoms (blue spheres) join together to form an inverted outer tetrahedron (OT) which is rotated by 30 degree with respect to the IT. (c) Six Mn atoms (magenta spheres) constituting an octahedron (OH) in a manner that makes each OH atom close to two IT and two OT atoms. (d) The IT, OT and OH atoms caged inside the cuboctahedron (CO) formed by 12 Mn atoms (orange spheres). Altogether, this arrangement of IT+OT+OH+CO forms cluster-A. (e) Cluster-B is formed in a similar way to that of cluster-A. Here the CO is formed by 11 In atoms (light green spheres) and one Mn atom (orange spheres),   the OH is composed of four Mn atoms (magenta spheres ) and two In atoms (light green spheres). The atomic composition/geometry for the IT and OT remains same as of cluster-A. (f) Extended view of the complete unit cell of Mn$ _3 $In with cluster-A  centered at (0, 0, 0) and cluster-B at (0.5, 0.5, 0.5).}
 	\end{center}
 \end{figure*}
\begin{figure}[tb!]
	\begin{center}
		\includegraphics[angle=0,width=8.5 cm,clip=true]{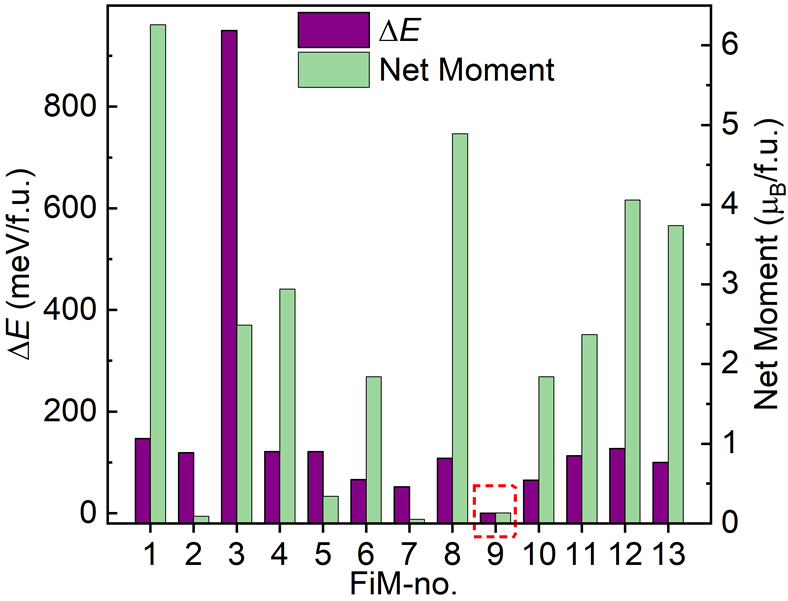}
		\caption{\label{FIG2} (Color online)  Relative formation energy ($\Delta$E) and net cell moments ($\mu$) of 13 different ferrimagnetic (FiM) spin configurations for pure Mn$ _{3} $In. The exact arrangement of spin alignments at different Mn-sublattices in each FiM configurations is shown in Table \ref{Configurations}. }
	\end{center}
\end{figure}

Although EB effects have been studied in few systems, the phenomenon is rarely observed in FCF materials. The EB phenomenon can be integrated  with the FCF to realize a thorough control over the interfacial exchange interactions. FCF are a unique class of materials where the properties of both antiferromagnets (e.g. vanishing net moment) and ferromagnets (e.g. large spin polarization) can be realized in a single system. The magnetic moment at different inequivalent sublattices adds up to give a vanishing net moment in FCF \cite{Coey14,14,Sahoo2016,Stinshoff2017,Giri2020}. The fully compensated magnetic state categorically differs from the AFM one, as the latter  possesses a magnetic inversion symmetry in contrast to the former.
Some of the unique properties and advantages of FCF materials are (i) the presence of nearly zero magnetic moment, which creates no external stray fields, (ii) spin sensitivity without stray magnetic fields, which allows them not to disturb the spin character and makes them ideal for spin-polarized scanning tunneling microscope tips and improved density of circuit integration in a chip, (iii) the low shape anisotropy, which helps in applications in spin injection, etc. In the present study, we have considered  a  special type of FCF system Mn$ _3 $In, where the magnetic compensation is achieved by AFM alignment of  two atomic clusters (as opposed to individual atomic moment in conventional AFM) in a single unit cell.  The presence of a long range FiM/AFM ordering with Neel temperature T$ _{N }$$\sim $75~K has been reported earlier \cite{17,Zhang2010}. We take advantage of the complex/unique atomic arrangement and the presence of intra-unit cell clusters to tune the EB in the system. Our assessment is based on the fact that the magnitude of antisite disorder can be readily modified by the substitution of non-magnetic/magnetic atoms, thereby, tailoring the long range magnetic order to formulate an inhomogeneous magnetic ground state. Interestingly, doping with non-magnetic heavy metals with strong spin-orbit coupling gives us the luxury to play with local crystalline symmetry to induce Dzyaloshinskii-Moriya interaction (DMI).  We propose a material platform [Pt and Ni doped cubic Mn$ _{3} $In alloys]  that can simultaneously host the existence of large EB and FCF behavior.

	
	


\begin{table}[ht]
	\begin{center}
		\begin{tabular}{|p{0.12\linewidth}|p{0.05\linewidth}|p{0.05\linewidth}|p{0.05\linewidth}|p{0.05\linewidth}|p{0.05\linewidth}|p{0.05\linewidth}|p{0.05\linewidth}|p{0.05\linewidth}|p{0.14\linewidth}|p{0.14\linewidth}|}
		\hline 
		Config. & 4e$_1$ & 4e$_2$ & 4e$_3$ & 4e$_4$ & 6f & 12i  & 12i & 6g &\centering$\mu$   &\centering $\triangle$E 
		\tabularnewline 
		& \centering A & \centering B& \centering A & \centering B & \centering A & \centering A & \centering B & \centering B &\centering in & \centering in
		\tabularnewline
		& \centering IT & \centering IT & \centering OT & \centering OT & \centering OH & \centering CO & \centering CO &  OH & \centering$\mu_B$/f.u.& \centering eV/f.u. \tabularnewline
		\hline 
		\hline 
		NM &- &- &- &- &- &- &- &- &- &1.68
		\tabularnewline
		\hline	
		FiM-1     &$\downarrow$  &$\uparrow$ &$\uparrow$ &$\uparrow$ & $\uparrow$&$\uparrow$ &$\uparrow$ &$\uparrow$ &6.26 &	0.147 \tabularnewline
		\hline
		FiM-2    &$\downarrow$  &$\downarrow$ &$\uparrow$ &$\uparrow$ &$\downarrow$ &$\uparrow$ &$\uparrow$ &$\downarrow$ &0.09  &0.119    \tabularnewline
		\hline 	 	
		FiM-3     &$\downarrow$  &$\uparrow$ &$\uparrow$ &$\downarrow$ &$\downarrow$ &$\uparrow$ &$\downarrow$ &$\uparrow$ &2.49  &0.95 \tabularnewline
		\hline
		FiM-4    &$\uparrow$ &$\uparrow$ &$\uparrow$ &$\uparrow$ &$\downarrow$ &$\downarrow$ &$\downarrow$ &$\downarrow$ &2.94 &0.121 \tabularnewline
		\hline
		FiM-5  &$\downarrow$ &$\downarrow$ &$\uparrow$ &$\uparrow$ &$\downarrow$ &$\uparrow$ &$\downarrow$ &$\downarrow$ &0.34  &0.121 \tabularnewline
		\hline 
		FiM-6  &$\uparrow$ &$\downarrow$ &$\downarrow$ &$\uparrow$ &$\downarrow$ &$\uparrow$ &$\uparrow$ &$\uparrow$ &1.84 &0.066 \tabularnewline
		\hline
		FiM-7  &$\downarrow$ &$\uparrow$ &$\uparrow$ &$\downarrow$ &$\uparrow$ &$\downarrow$ &$\downarrow$ &$\uparrow$ &0.05  &0.052  \tabularnewline
		\hline
		FiM-8  &$\uparrow$ &$\uparrow$ &$\downarrow$ &$\downarrow$ &$\downarrow$ &$\downarrow$ &$\uparrow$ &$\downarrow$ &4.89 &0.108 \tabularnewline
		\hline 	
		\textbf{FiM-9}  &$\uparrow$ &$\uparrow$ &$\downarrow$ &$\downarrow$ &$\downarrow$ &$\uparrow$ &$\uparrow$ &$\downarrow$ &\textbf{0.13}  &\textbf{0} \tabularnewline
		\hline
		FiM-10 &$\downarrow$ &$\uparrow$ &$\uparrow$ &$\downarrow$ &$\uparrow$ &$\downarrow$ &$\downarrow$ &$\downarrow$ &1.84  &0.065 \tabularnewline
		\hline
		FiM-11 &$\uparrow$ &$\uparrow$ &$\downarrow$ &$\downarrow$ &$\uparrow$ &$\downarrow$ &$\downarrow$ &$\downarrow$ &2.37  &0.113 \tabularnewline
		\hline
		FiM-12 &$\downarrow$ &$\uparrow$ &$\uparrow$ &$\downarrow$ &$\downarrow$ &$\downarrow$ &$\downarrow$ &$\downarrow$ &4.06 &0.127 \tabularnewline
		\hline
		FiM-13 &$\downarrow$ &$\uparrow$ &$\uparrow$ &$\downarrow$ &$\downarrow$ &$\downarrow$ &$\uparrow$ &$\downarrow$ &3.74 &0.100 \tabularnewline
		\hline   
		\end{tabular}
	\end{center}	

	\caption{\label{tab1}Total cell moments ($\mu$) and relative energies ($\Delta$E) of $13$ different ferrimagnetic(FiM) spin configurations for pure Mn$_3$In. FiM-9 is the lowest energy configuration which is set as the reference configuration with energy zero.   4e$_1$, 4e$_2$, 4e$_3$, 4e$_4$, 6f, 12i and 6g are the seven inequivalent Wyckoff positions for Mn in pure Mn$_3$In. NM = Non-magnetic configuration. }
	\label{Configurations}
\end{table}

 \section{\textbf{Methods}}
 
 Density functional theory (DFT) \cite{s1} calculation was carried out using Viennna $ ab initio $ simulation pPackage (VASP) \cite{s2,s3,s4} with a projected augmented wave basis \cite{s5} and the generalized gradient approximated (GGA) exchange-correlation functional of Perdew-Burke-Ernzerhof (PBE) \cite{s7}. A plane wave energy 
 cutoff of 400 eV was used. The Brillouin zone integration was done using a  6$\times$6$\times$6 $\Gamma$-centered k-mesh.  Polycrystalline ingots of Mn$_ {3-x} $Pt$_x $In for $ x = $ 0.1, 0.2 and 0.3, and Mn$ _{3-y} $Ni$_y $In with $ y = $ 0.1 and 0.2 were prepared by arc-melting technique.  The appropriate ratio of the respective elements were taken and melt under Ar gas atmosphere within the arc melt chamber. As prepared samples were enclosed within a quartz tube under Ar atmosphere. After that, a heat treatment for 8 days at 1073 K temperature was completed and subsequently quenched in the ice water mixture. Room temperature x-ray powder diffraction measurements were performed using a Rigaku SmartLab x-ray diffractometer with a Cu-K$_\alpha  $ source to characterize the structural phase. To probe the compositional homogeneity, field emission scanning electron microscope (FESEM) equipped with energy dispersive x-ray (EDX) analysis was utilized. Low field magnetic measurements were carried out using SQUID vibrating sample magnetometer (MPMS-3, Quantum Design) and VSM option in Quantum Design physical property measurement system (PPMS). Transport measurements were performed using Quantum Design PPMS. Pulsed field magnetization measurements up to 60 T were carried out at the Dresden High Magnetic Field Laboratory HLD-HZDR. The 30 T static magnetic field measurements were performed using a vibrating sample magnetometer at the High Field Magnet Laboratory HFML-RU/FOM in Nijmegen. 

 \section{\textbf{THEORETICAL CALCULATION}}
 Mn$ _{3} $In crystallizes in a cubic structure with 52 atoms in a unit cell, which comprises of two atomic clusters, each containing 26 atoms and centered at (0, 0, 0) and (0.5, 0.5, 0.5) \cite{19}. To understand the exact nature of atomic arrangement we have decomposed the crystal structure based on the general symmetry analysis as depicted in Fig. 1(a)-(f). Each cluster (labeled A and B) consists of an inner tetrahedron (IT), outer tetrahedron (OT), octahedron (OH) and cubo-octahedron (CO). In case of cluster-A, all the sites are occupied by Mn atoms only [Fig. 1(d)]. For cluster-B, IT and OT are fully occupied by Mn atoms, while OH and CO positions are mostly filled by the Mn and In atoms, respectively (with OH containing two In and rest Mn while CO invloves one Mn and rest In atoms), as shown in Fig. 1(e). The mixed occupancy between Mn and In atoms at the OH and CO sites of cluster-B intrinsically induces antisite disorder in the system. In our study, we mostly concentrate on the Ni and Pt doped Mn$ _{3} $In alloys to facilitate tunable magnetic ordering. 
 \begin{table}[t]
 	\begin{center}
 		
 		\begin{tabular}{|p{0.11\linewidth}|p{0.06\linewidth}|p{0.06\linewidth}|p{0.07\linewidth}|p{0.07\linewidth}|p{0.07\linewidth}|p{0.07\linewidth}|p{0.06\linewidth}|p{0.06\linewidth}|p{0.07\linewidth}|p{0.08\linewidth}|p{0.07\linewidth}|}
 			\hline 
 			Conc. & 4e$_1$ & 4e$_2$ & 4e$_3$ & 4e$_4$ & 6f & 6g& 12i  & 12i  &$\mu_{net}^{(A)}$
 			&$\mu_{net}^{(B)}$
 			
 			& $\mu_{eff}$
 			\tabularnewline
 			&\centering A &\centering B &\centering A&\centering B &\centering A &\centering B &\centering A& \centering B& & &
 			
 			\tabularnewline
 			\hline 
 			\hline 
 			Mn$_{3}$In  &0.6  &1.7 &-2.4 &-2.4 & -2.8&-3.5& 3.3&3.6 &14.8 &-13.4& 0.1
 			\tabularnewline
 			\hline	
 			$x$=0.15     &0.6 &1.8 &-2.4 &-2.6 & -2.9 &-2.7& 2.9&3.6 & 11.4&	-10.2 & 0.1 \tabularnewline
 			\hline
 			$y$=0.15 &0.6 &1.8 &-2.4 &-2.9 & -2.7 &-2.7& 2.9&3.6 & 12.3 & -11.2& 0.1 \tabularnewline
 			\hline 	 	
 			$x$=0.3 &0.6 &2.0 &-2.5 &-2.6 & -2.9 &-1.8& 2.7&3.7& 8.1 & -5.5& 0.2 \tabularnewline
 			\hline 
 			$y$=0.3   &0.6 &2.0 &-2.5 &-2.6 & -2.9 &-1.7& 2.7&3.7  & 7.2 & -5.6& 0.1 \tabularnewline
 			\hline 
 		\end{tabular}
 	\end{center}
 	\caption{ The moments (in $\mu_B$) at different Wyckoff positions in cluster-A and B of Mn$ _{3} $In, Mn$_{3-x} $Pt$ _{x} $In, and Mn$_{3-y}$Ni$_{y}$In.  The net moments of cluster-A($\mu_{net}^{(A)}$)  and cluster-B($\mu_{net}^{(B)}$) are coupled antiferromagnetically. The effective moment ($\mu_{eff}=\mu_{net}^{(A)}+\mu_{net}^{(B)}$) is in $\mu_B$/f.u. The individual site moments are rounded up to the first decimal place, while the net cluster moments are calculated taking the second decimal place into consideration. The effective moment is also rounded up to the first decimal place.} 
 	\label{Configurations1}
 \end{table}

 \begin{figure}[tb!]
 	\centering
 	\includegraphics[angle=0,width=8.5 cm, clip=true]{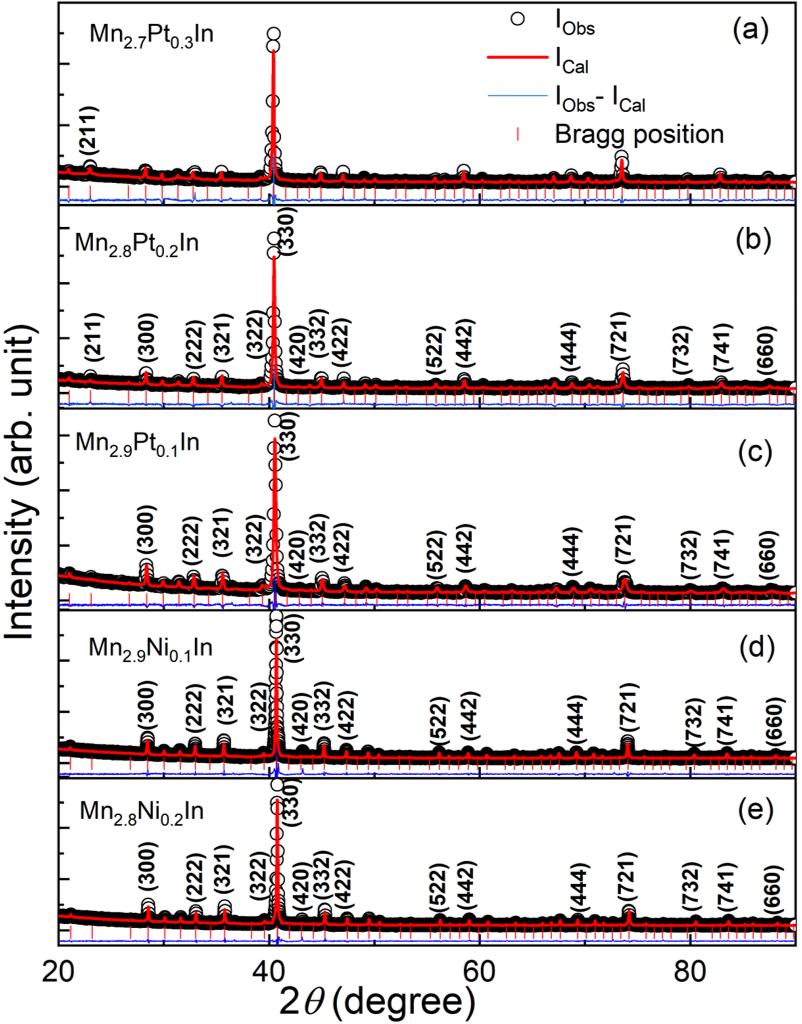}
 	\caption{ Room temperature powder x-ray diffraction (XRD) pattern along with Reitveld refinement for, (a)-(c) Mn$_{3-x}$Pt$_{x}$In ($ x= $0.1 to 0.3) and (d), (e) Mn$_{3-y}$Ni$_{y}$In (y=0.1 and 0.2) .The experimentally observed intensity  (I$ _{Obs} $) and simulated intensity (I$ _{Cal} $) data are represented by the black open circles and red solid lines, respectively. The solid lines in blue color describe the difference between the experimental and simulated intensity. The red vertical lines indicate the Bragg positions.}
 	\label{figure3}
 \end{figure}
 

 To understand the magnetic ground state of the present systems, we have simulated different spin configurations by varying the spin alignment at various Wyckoff positions of the Mn sublattices within density functional theory (DFT). Figure 2 shows the relative formation energies ($\Delta$E) and net cell moments ($\mu$) of 13 different spin configurations (labelled FiM). The structures of each of these configurations were fully optimized to reach their minimum energy. As evident, FiM-9 turns out to be energetically the most favourable, with negligibly small net moment.  The relaxed lattice constant for this configuration is 9.25\r{A}. The exact form of the 13 different spin configurations and their respective $\Delta$E and $\mu$ are given in Table \ref{Configurations}. A perfect FM phase could not be stabilized \cite{FM_discuss}. Keeping in mind the complex nature of magnetic ordering in Mn$ _{3} $In, the degree of frustration around each Mn-sites is quite obvious. We have calculated the total degree of frustration (summing up the frustration at each inequivalent Mn-sites) for all the 13 configurations (see Fig. 18 within the Supplemental Material  \cite{supplementary} for details).

The magnetic ground state for the  Pt and Ni doped samples are calculated by considering  energetically the  most stable Mn$_3$In configuration (FiM-9) and then simulating the magnetic structure for Mn$_{2.85}$(Pt, Ni)$ _{0.15} $In and Mn$_{2.7}$(Pt, Ni)$ _{0.3} $In. The resulting net cell moments are listed in Table \ref{Configurations1} . It is evident that the substitution of Pt in Mn$_3$In does not alter the fully compensated ferrimagnetic behavior of the host material.  Most importantly,  for all the cases, each individual cluster consists of a large net magnetic moment. The net magnetic moment in cluster-A is almost equal and opposite to that of cluster-B, resulting in a nearly fully compensated magnetic state. One can notice that although FiM-9, with a fully compensated magnetic configuration, is the ground state, a net uncompensated magnetic moment of 1.84 $\mu_B$/f.u. can also be found in case of FiM-6 and FiM-10 arrangements with a very small energy difference  and almost same degree of frustration to that of FiM-9. Hence, it might be possible to stabilize  some uncompensated FiM magnetic clusters within the fully compensated ferrimagnetic host. Therefore, the exchange interaction between the fully compensated FiM host and the uncompensated magnetic clusters can give rise to possible EB effect.
\section{\textbf{STRUCTURAL ANALYSIS}} 
To verify our theoretical propositions, we have synthesized Mn$ _{3-x}$Pt$_{x} $In and Mn$ _{3-y}$Ni$_{y} $In samples with $ x=  $ 0.1 to 0.3 and $ y= $0.1 to 0.2.  We only concentrate on the doped samples as it is not possible to stabilize a single structural phase of Mn$ _{3} $In by the present arc-melting technique. The structural phase purity of all the doped samples can be seen from the Rietveld refinement of the room temperature powder x-ray diffraction (XRD) data as depicted in Fig. 3.  All the Bragg peaks observed experimentally can be well indexed by incorporating the structural symmetry associated with space group P$ \bar{4} $3m. The lattice parameters and the other agreement factors obtained from Rietveld refinement are tabulated within the Supplemental Material \cite{supplementary}. We have also thoroughly investigated the XRD data to find any site-specific preference of the doping element. As can be seen from the Pt composition-dependent room temperature powder XRD patterns [Fig. 3(a)-3(c)], the Bragg peak (211) is absent in Mn$ _{2.9} $Pt$ _{0.1} $In [Fig. 3(c)]. The intensity of the Bragg peak (211) increased with the increase of Pt concentration, in fact, it is very prominent in Mn$ _{2.7} $Pt$ _{0.3} $In [Fig. 3(a)]. Therefore, we have systematically substituted Pt atoms in all the possible Wyckoff positions to find out any preferentially site occupation. The variation of simulated intensity of the (211) peak for substitution of Pt at different Wyckoff positions in the Mn$_{2.7}$Pt$_{0.3}$In is shown in Supplemental Material \cite{supplementary}. In particular, we find a site preferential occupancy of doped Pt, Ni atoms at AOH and BOH sites (see Supplemental Material)\cite{supplementary}.
\begin{figure*}[tb!]
	\begin{center}
		\includegraphics[angle=0, width=17 cm,clip=true]{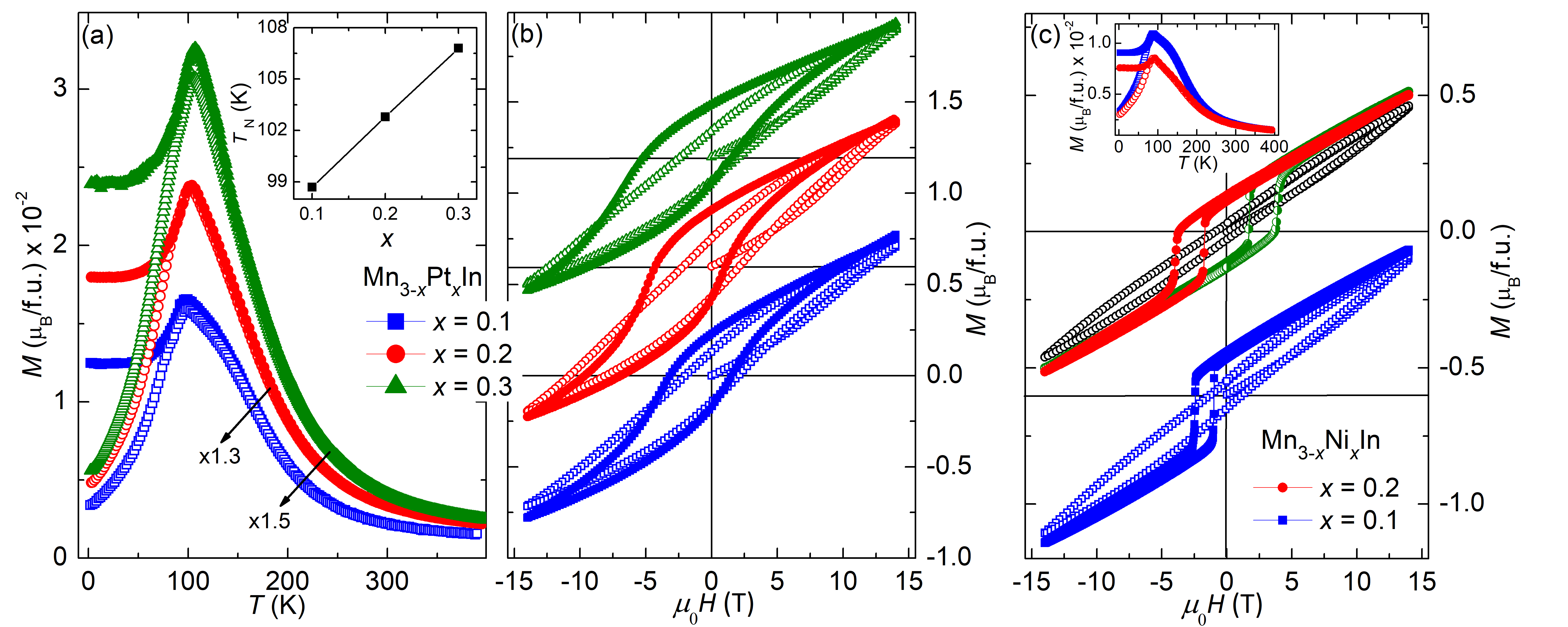}
		\caption{\label{Figuremerged} (a) Temperature dependent magnetization, $M(T)$, measured at 0.1~T for Mn$ _{3-x} $Pt$_{ x }$In. Open (solid) symbols correspond to the ZFC (FC) data, respectively. The data for $x=$0.2 and $x=$ 0.3 are multiplied by scaling factors of  1.3 and 1.5, respectively, for better visualization. Inset shows the Neel temperature (T$_{N }$) vs. $x$. (b) ZFC (open symbol) and +5~T FC (solid symbol) $M(H)$ loops measured  at $ T= $ 2 K.  $ M(H) $ loop corresponding to the sample $ x=  $0.2 and 0.3 are shifted by 0.6 and 1.2 $ \mu $$ _{B} $/f.u., respectively, along the magnetization axis. (c) ZFC (open symbol), +5 T FC (solid symbol) and -5 T FC (half filled symbol) $ M(H) $ loops for Mn$ _{3-y} $Ni$_{ y }$In measured at $ T= $ 2 K.  $M(H)$ loop for $ y= $ 0.1 is shifted by -0.6 $ \mu $$ _{B} $/f.u. The inset of (c) shows the \textit{M}(T) curve, squares and circles represent data for y = 0.1 and y = 0.2, respectively. } 
		
	\end{center}
\end{figure*}

\section{\textbf{MAGNETIZATION STUDY}}

To study the effect of Pt doping on the magnetic properties of Mn$_3$In, we have carried out temperature ($ T $) dependent magnetization  measurements  for the Mn$_{3-x}$Pt$_{x}$In samples, as shown in Fig. 4(a). All the samples exhibit a typical AFM type $M(T)$ curves. However, the  zero field cooled (ZFC) and field cooled (FC) $M(T)$ curves display a strong  bifurcation below the  Neel temperature ($T_N $), which increases monotonically with increasing $x$ [inset of Fig. 4(a)].  The low temperature irreversibility between the ZFC and  FC $M(T)$ data suggests the presence of a magnetic phase coexistence, which might be due to the formation of FM cluster within the compensated FiM matrix. The presence of large irreversibility between the ZFC and FC $M(T)$ data even at  5 T and frequency dependent ac-susceptibility measurements  eliminate the possible existence of SG phase in the present system (see Supplemental Material)\cite{supplementary}.  


\begin{figure}[tb!]
	\centering
	\includegraphics[angle=0,width=8 cm, clip=true]{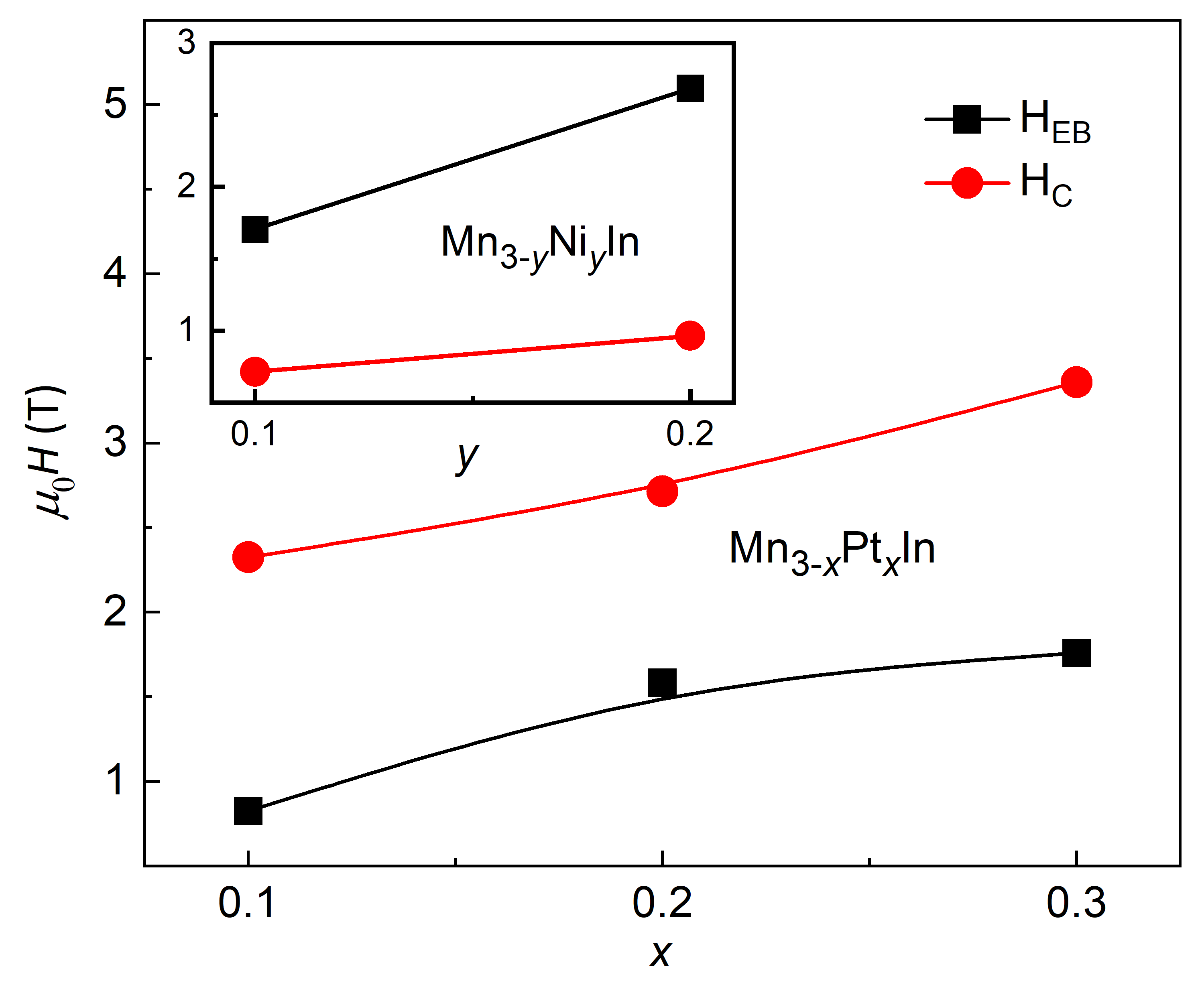}
	\caption{ Pt concentration (x) dependent exchange bias fields ($ H_{EB} $) and coercive fields $ H_C $ for Mn$ _{3-x} $Pt$_{ x }$In. Inset shows $ H_{EB} $ and $ H_C $ for Mn$ _{3-y} $Ni$_{ y }$In. }
	\label{figure6}
\end{figure}

In order to further examine the nature of the magnetic phase coexistence, we have performed ZFC isothermal magnetization measurements  at 2 K, as shown in Fig. 4(b) (open symbols).  The ZFC $ M(H) $ loops exhibit a nearly linear field dependency up to the measured field of 14 T. However, all the loops show hysteretic behavior with a large coercive field ($ H_{C} $). The results corroborate our assumption about the presence of mixed FM and FiM phases. The linear nature of $ M(H) $ loop may arise from  the compensated FiM background, whereas,  the hysteretic behavior can originate from the existence of FM clusters.   We have further measured FC isothermal magnetization loops at 2~K to examine the presence of any exchange interaction between the FM and FiM phases, as shown in Fig. 4(b) (solid symbols). Before the measurement, the samples were cooled down to the required temperature  from 300~K in presence of 5~T field. As it can be seen, all the FC $ M(H) $ loops display a large shift along the negative field axis, demonstrating the existence of a large unidirectional exchange anisotropy in the system.  Like the Pt doped samples, the $ M(T) $ data for Mn$_{3-y}$Ni$_{y}$In also displays a large irreversible behavior [inset of Fig. 4(c)]. The ZFC and FC $ M(H) $ measurements at 2 K for the Ni doped samples are plotted in Fig. 4(c). The ZFC $ M(H) $ loops exhibit a similar kind of hysteretic behavior as that of Pt doped samples, whereas, the FC loops display a spontaneous magnetization behavior with the loop closing  field of about 5~T. The EB and coercive fields are calculated by using the formula $H_{EB}$ =-($H _{L} $+$H _{R} $)/2 and $H_{C}$= $ \mid $ $H_{L} $-$H _{R} $ $ \mid $/2 respectively, where  $H _{L} $ and $H _{R} $ are the lower and upper critical field at which the magnetization becomes zero. We find a large EB field ($H_{EB}$) of 0.8, 1.6, and 1.8 T for $ x= $0.1, 0.2, and 0.3, respectively [Fig. 5]. In addition,  we also observe an enhancement in the $H_{C}$ values to  2.33 T, 2.67 T and 3.37~T for  $ x= $0.1, 0.2, and 0.3, respectively. For the Ni doped samples, an extremely large $H_{EB}$ of 2.68 T is found for Mn$_{2.8}$Ni$_{0.2}$In [see inset of Fig. 5]. 


\begin{figure}[tb!]
	\centering
	\includegraphics[angle=0,width=8.5 cm, clip=true]{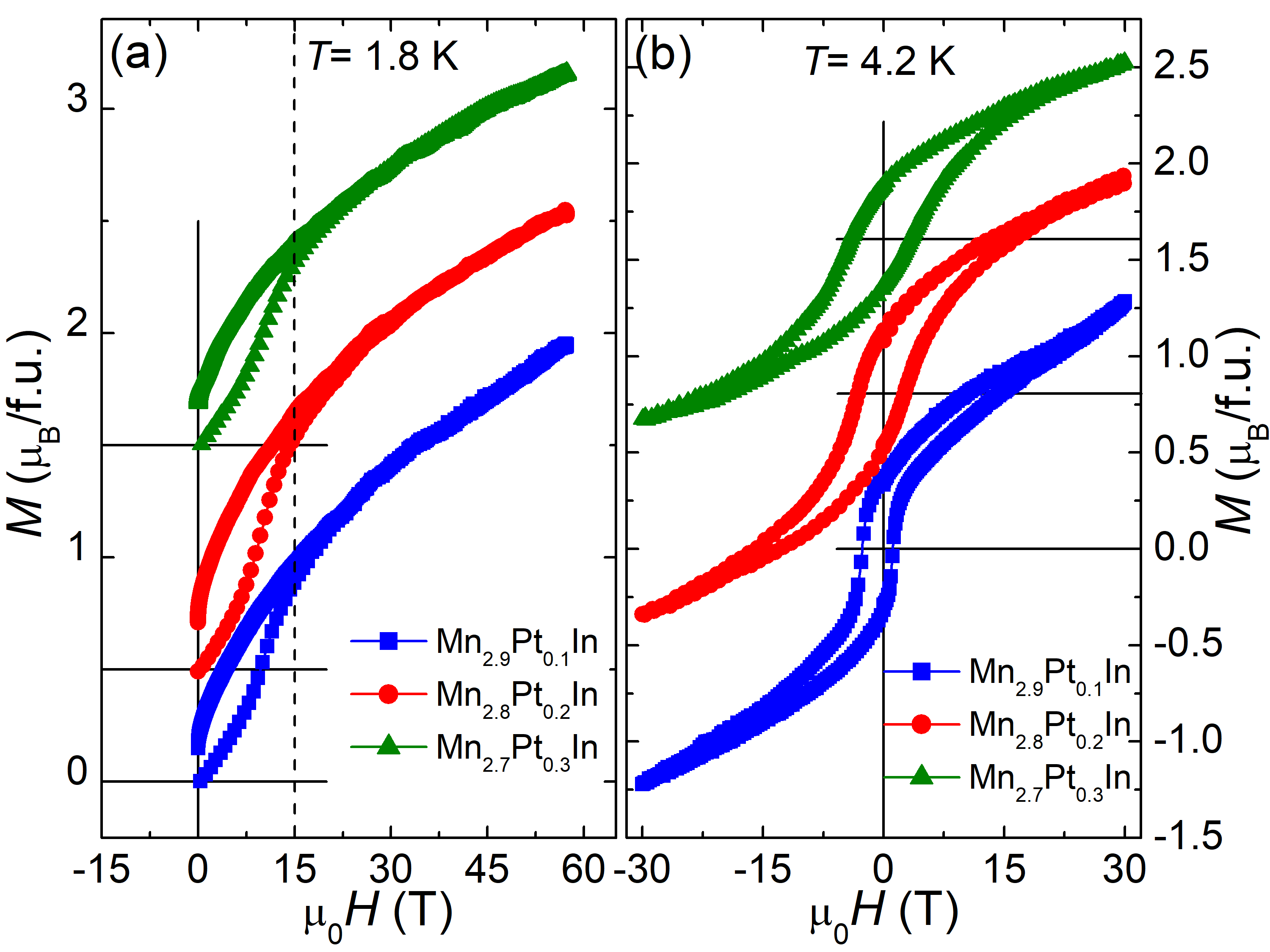}
	\caption{ (a) ZFC $ M(H) $ loops measured up to 60 T for Mn$_{3-x}$Pt$_{x}$In. Magnetization data for the sample $ x= $ 0.2 and 0.3 are shifted by 0.5 and 1.5 $ \mu $$ _{B} $/f.u., repectively along $ M $-axis. (b) $ M(H) $  loops measured up to $\pm 30$ T after field cooling the sample in an applied field of $H_{FC}$ = 15 T. $ M(H) $ loops of $  x= $ 0.2 and 0.3 are shifted by 0.8 and 1.6 $ \mu $$ _{B} $/f.u., respectively.}
	\label{figure7}
\end{figure}

\begin{figure*} [t]
	\begin{center}	
		\includegraphics[angle=0,width=14 cm, clip=true]{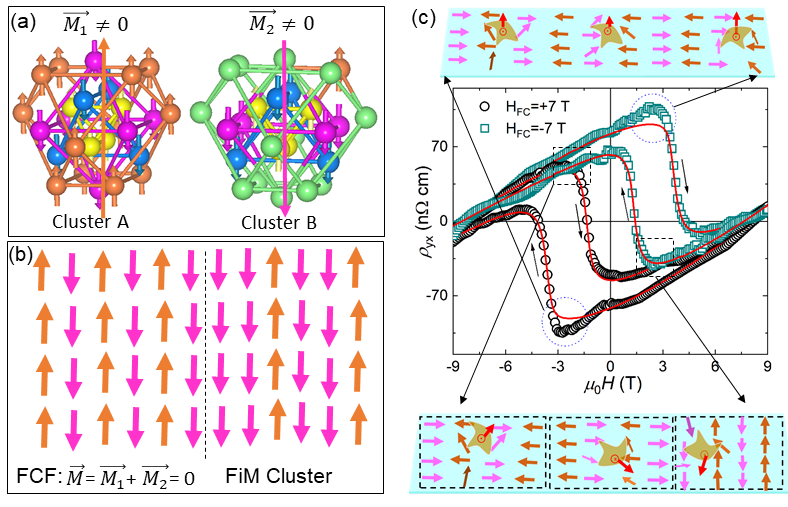}
		\caption{\label{FIG4}(Color online)  (a) Cluster-A and cluster-B with finite staggered magnetization aligned antiparallely. (b) Left panel: Fully compensated FiM background as a result of cancellation between the staggered magnetization of cluster-A (orange arrows)and cluster-B (magenta arrows). Right panel: Possible FiM cluster due to a finite net (cell) magnetic moment, as in FiM-6/FiM-10 configuration in Table \ref{Configurations}. The dashed line indicates the interface between the FCF background and the FiM cluster with finite moment. (c) Field dependence of Hall resistivity ($\rho_{yx}$) measured at 5~K after field cooling the sample in an applied field of +7~T (open circles) and - 7~T (open squares) for Mn$_{2.8}$Ni$_{0.2}$In. The open symbols represent experimental data while the solid lines correspond to the total calculated Hall resistivity. The schematic spin configuration at the top and bottom panels represent the possible interfacial FCF magnetic configuration at the field reversal point which are marked in dotted circles and dotted squares, respectively.  The top and bottom panel depicts the subtended solid angle by the noncoplanar spin contain with in the single and multidomain AFM, rspectively.The solid angle subtended by the non-coplanar spins in the schematic diagrams are marked by shaded dark yellow region and the red arrows indicate the direction of fictitious magnetic field.}
	\end{center}
\end{figure*}



The finding of large EB  with contrasting nature of the FC hysteresis in Pt and Ni doped samples require a deeper understanding of the underlying mechanism. Although the FC loops for the Ni doped samples closes at a moderate field of about 5~T, it is unclear for the Pt doped samples, if the FC $ M(H) $ loops really closes at 14~T. Moreover, to make sure that the observed EB effect  does not arise from the minor loop measurement, we have carried out pulsed field magnetization measurements at 1.8 K at a field up to 60 T for the Pt doped samples as shown in Fig. 6(a). Clearly, all the loops close at a field of about  $  \sim $15 T. The $M(H)$ curves do not show any signature of saturation up to the maximum applied field, indicating the robustness of inter sublattice exchange strength in the compensated ferrimagnet. A strong hysteretic behavior in the field range of 0-15 T with a remanent magnetization of 0.16-0.23 $ \mu $$ _{B} $/f.u. is also found in all the three samples. In addition, the presence of a  step like feature around 10~T also suggests a field induced metamagnetic type of transition. To further probe the effect of high magnetic field, we have taken 15~T FC hysteresis loops measured up to 30~T for the Pt doped samples, as depicted in Fig. 6(b). As evident, all the $ M(H) $ loops close at about 15~T (see ZFC 60~T measurements). Hence, the FC hysteresis loops measured up to a field of 14~T in Fig. 4(b) fall almost in the major loop category. However, we find that  the 15~T FC $ M(H) $ loops (measured  up to 30~T) exhibit smaller EB field in comparison to that of the 5~T FC loops shown earlier.
 
\section{\textbf{TRANSPORT MEASUREMENT AND DISCUSSIONS}}
Obsevation of large EB in the present system strongly suggests the existence of exchange coupling between the FCF background [Fig. 7(a) and left panel of Fig. 7(b)] and the FiM clusters [right panel of Fig. 7(b)]. Our experimental results also directly support the  theoretical proposition of FCF  as the lowest energy state with negligible energy difference between the FCF and uncompensated FiM states. In a recent theoretical study, it is proposed that the uncompensated ferrimagnetic ordering as the lowest energy state in Mn$ _{3} $In \cite{Chatterjee2020}. However, our theoretical as well as the experimental results categorically establish FCF as the magnetic ground state in the present system. In order to further understand the nature of exchange interactions between the FCF background and the FiM cluster, we have performed Hall measurements in Mn$_{2.8}$Ni$_{0.2}$In, as it requires a moderately low field ($\sim$ 5 T) to close the hysteresis loop. The field dependence of Hall resitivity ($ \rho_{yx} $) measured after field cooling the sample in +7~T and -7~T is shown in Fig. 7(c). To extract any possible additional component in the Hall resistivity, we have fitted the experimental $ \rho_{yx} $ data with the calculated one. The total $ \rho_{yx} $ can be expressed as $\rho_{yx} = \rho_{N}+\rho_{AH}+\rho^{EH}_{yx}$, where $\rho_{N}$, $\rho_{AH}$ and $\rho^{EH}_{yx}$ are  normal, anomalous and extra Hall resistivities, respectively \cite{Giri2020, Rout19}. $\rho_N$ can be written as, $\rho_{N} = R_{0}H$, where $R_0$ is the normal Hall coefficient and $ H $ is the magnetic field. While $\rho_{AH}$ can be expressed as,  $\rho_{AH} = b\rho_{xx}^2 M$, where $b$ is a constant, $\rho_{xx}$ is the longitudinal resistivity and $ M $ is the magnetization. Since the FC loop closes for field above 5~T, it can be assumed that the spin structure saturates for field larger than 5~T and hence the high field $\rho_{yx}$ data only consists of  $\rho_{N}$ and $\rho_{AH}$. In this scenario, the $\rho_{yx}$ at field greater than 5~T can be expressed as $\rho_{yx} = R_{0}H + b\rho_{xx}^2 M$. The calculated Hall resistivity is plotted as solid lines in Fig. 7(c). If we see the +7~T FC data, the experimental and calculated curves match pretty well everywhere except at the  magnetization reversal point at +9~T to -9~T curve. To further examine this unusual behavior, we have performed similar fitting for the -7~T FC data, where the difference between the  experimental and the calculated data arises at the magnetization reversal point at -9~T to +9~T field sweep curves. This indicates the presence of extra Hall effect (EHE) at the field reversal marked by dotted circles in Fig. 7(c). It is to be noted that for the +7~T FC loop, the measurement was performed by sweeping the field from +9 $\rightarrow$ -9 $\rightarrow$ +9~T, whereas, the field was sweeped as -9 $\rightarrow$ +9 $\rightarrow$ -9~T for -7~T FC loop. We have also performed the Hall effect measurements by zero field cooling the sample to low temperature. As expected, we do not find any extra Hall component in case of the ZFC Hall measurements (see Supplemental Material) \cite{supplementary}.

The observed  differences between the calculated and experimental Hall resistivity curves indeed indicate towards a different mechanism of magnetization reversal through lower and upper critical field in an exchange coupled system. The often found asymmetry in hysteresis curve in the vicinity of magnetization reversal occurs due to either domain wall motion or magnetization rotation on opposite side of hysteresis \cite{PRL2000,McCord2003,Brems2005}. Most importantly, this extra Hall contribution in the present system appears only in the field decreasing or field increasing path of the Hall measurement when the sample is field cooled in positive or negative field. This kind of extra Hall contribution can only be assigned to the interfacial non-coplanar spin structure in an exchange bias system. This is due to the fact that the sign of the cooling field determines the nature of the spin orientations at the interface, leading to the observed asymmetry in the Hall signal across the field -reversal regime. Similar kind of effect has also been previously found in an exchange
bias system, where, the additional Hall contribution is attributed to the topological -spin texture at the interface\cite{Meng2008}. Therefore, in the present system, the origin of EHE is most likely connected to the non-vanishing scalar spin chirality originated from the non-coplanar spin structure\cite{Surger2014, Wang2019,Giri2020,Rout19}. As a result, an extra component of Hall resistivity appears in the vicinity of the magnetization reversal through magnetization rotation [marked in dotted circles in Fig. 7(c)]. A possible spin configuration corresponding to the Hall signal marked in dotted circles is illustrated in the top panel of Fig. 7(c). In this scenario, the fictitious magnetic fields (indicated by the red arrows) associated with the solid angle subtended by the non-coplanar spins within the single domain FCF add up to give a non-vanishing effective field. This gives rise to the observed additional Hall component as marked by dotted circles in Fig. 7(c). On the other hand, the absence of extra Hall component in the reverse cycle [marked as dotted squares
and represented schematically in the bottom panel of Fig. 7(c)] indicates the formation of multidomain state in the FCF layer. The random orientation of the fictitious magnetic field in different domains leads to a vanishing effective field. Moreover, the nucleation of domain states is favourable while moving from negative saturated field to positive one for the positive field cooled case and vice versa. In fact, the existence of AFM domain at the interface in an EB system has also been reported earlier \cite{McCord2003,Nat2000,Ohldag2001}.

\section{\textbf{Conclusions}}
In conclusion, we have reported a classic example of a composite quantum material where the material (Mn$_3$In) shows the co-existence of fully compensated ferrimagnet (FCF) and large exchange bias (EB). In both pure and undoped Mn$_3$In, FCF arises from the antiferromagnetic coupling between the intra-cluster staggered moment. The high degree of frustration in the energetically most stable FCF state and small energy barrier of this state with respect to other uncompensated magnetic states are the most plausible reason to give rise to a magnetic inhomogeneous state. The finding of large exchange bias in the present system is the  manifestation of the exchange interaction between the FCF and the uncompensated ferrimagnetic clusters. The observed EHE in such systems indicate presence of interfacial DMI along with the symmetric exchange interaction. Moreover, it establishes the importance of compensation to achieve large EB  regardless of the crystalline anisotropy.



\begin{acknowledgments}
	
	AKN acknowledges the support from Department of Atomic Energy (DAE), the Department of Science and Technology (DST)-Ramanujan research grant (No. SB/S2/RJN-081/2016), SERB research grant (ECR/2017/000854) and Nanomission research grant  [SR/NM/NS-1036/2017(G)] of the Government of India. AA acknowledges DST-SERB (Grant No. CRG/2019/002050) for funding to support this research. AA thank A.I. Mallick for some initial calculations and discussion. We acknowledge the support of HLD at HZDR and HFML, members of the European Magnetic Field Laboratory (EMFL).

\end{acknowledgments}
\bibliographystyle{MSP}

\begin{thebibliography}{100}
\bibitem{FM_discuss}{Interestingly, it is not possible to stabilize a ferromagnetic (FM) configuration for Mn3In. In fact, the FiM-1 calculation was initiated with a perfect FM configuration to start with, but the spin at 4e1 Mn-sites (inner tetrahe- dron of cluster A) flip after the relaxation, and converged to FiM-1.}
\end{thebibliography}

\section*{Supplemental Information}
\section{Characterization}

\subsection{XRD Analysis}

\begin{figure} [t]
	\begin{center}	
		\includegraphics[angle=0,width=8 cm, clip=true]{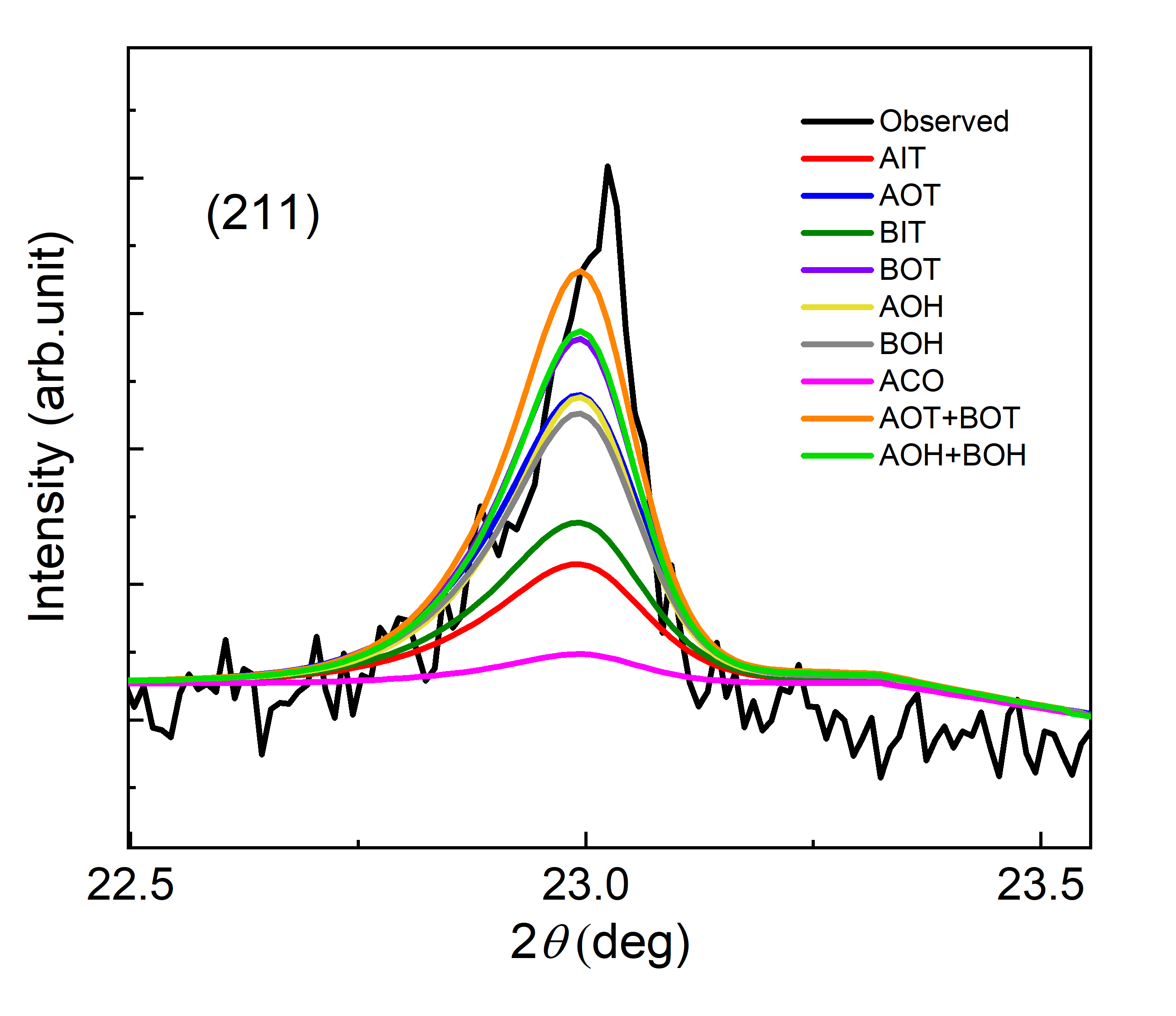}
		\caption{\label{FIG8}(Color online) The variation of the (211) peak intensity with the different combination of Wyckoff position of Pt atoms in Mn$ _{2.7} $Pt$ _{0.3}$In}
	\end{center}
\end{figure}
We find that it is not possible to stabilize a single structural phase of Mn$ _{3}$In by the present arc-melting technique. However, we were able to prepare Mn$ _{3-x} $Pt$ _{x} $In and Mn$ _{3-y} $Ni$ _{y} $In (x=0.1 to 0.3 for Pt and y=0.1 to 0.2 for Ni) samples in a structurally single phase of $ \gamma $-brass structure.
To obtain the phase purity and structural information of Mn$ _{3-x} $Pt$ _{x} $In and Mn$ _{3-y} $Ni$ _{y} $In, we have performed the powder XRD measurement at room temperature. Reitveld refinement of the powder XRD data confirms that all samples are single phase.  It is found that the lattice constant increases/decreases with the increasing Pt/Ni doping concentration [see Table-\ref{Configurations3}]. We have thoroughly investigated the XRD data to find any site-specific preference of the doping element. As can be seen from the Pt composition-dependent room temperature powder XRD pattern [Fig. 3 of main text], the Bragg peak (211) is absent in Mn$ _{2.9} $Pt$ _{0.1} $In. With the increase of Pt concentration, the intensity of the Bragg peak (211) is also increased, in fact, it is very prominent in Mn$ _{2.7} $Pt$ _{0.3}$In. There are multiple distinguishable Mn positions within Mn$ _{3} $In. So, the Pt can go to at any position. However, we have tried to obtain a particular or combination of distinguishable Mn positions to be preferentially occupied by Pt atoms. Therefore, we have systematically substituted Pt atoms among all the possible Wyckoff positions. The simulated intensity of the (211) peak by substitution of Pt at different Wyckoff positions in the Mn$ _{2.7} $Pt$ _{0.3}$In has shown in Fig. 8. We have taken the simulated intensity of the (211) peak, reduced $ \chi  $$ ^2 $value, R$ _{p} $, and R$ _{wp} $ factors as the parameters to determine the possible favorable positions of Pt.  It is evident from the intensity variation with Pt position that the 2 Pt atoms at AOH and 2 Pt atoms at BOH position appear to produce the best agreement factor. Also, the equal distribution of Pt atom at AOT and BOT positions gives almost the same agreement factors which have been compared in Table-\ref{Configurations4}. We have tried other combinations as well, which do not agree well with the experimental data. Therefore, we assert that the Pt atoms are most favorably occupying the AOH and BOH positions. During the theoretical calculation, we have taken care of the site preferences of Pt atoms.
\begin{table}[t]
	\begin{center}
		
		\begin{tabular}{|p{0.25\linewidth}|p{0.15\linewidth}|p{0.15\linewidth}|p{0.15\linewidth}|p{0.15\linewidth}|}
			\hline 
			Composition & Lattice constant ($ \AA $) &$ \chi $$ ^2 $ & R$ _{p} $& R$ _{wp} $  
			\tabularnewline
		    \hline 
			Mn$ _{2.9} $Pt$ _{0.1}$In  &9.431  &1.87 &3.04&4.22
			\tabularnewline
			\hline	
		    Mn$ _{2.8} $Pt$ _{0.2}$In &9.449 &2.9 &4.83 &6.36 
		    \tabularnewline
			\hline
			Mn$ _{2.7} $Pt$ _{0.3}$In & 9.460 &3.85 &5.23 &7.49 
			\tabularnewline
			\hline 	 	
			Mn$ _{2.9} $Ni$ _{0.1}$In&9.399 &5.11&3.65 &5.61
			\tabularnewline
			\hline 
			Mn$ _{2.8} $Ni$ _{0.2}$In   &9.386 &3.68 &3.22 &4.73
			\tabularnewline
			\hline 
		\end{tabular}
	\end{center}
	\caption{ Comparison of lattice parameters and various agreement factors.} 
	\label{Configurations3}
\end{table}

\begin{table}[t]
	\begin{center}
		
		\begin{tabular}{|p{0.25\linewidth}|p{0.15\linewidth}|p{0.15\linewidth}|p{0.15\linewidth}|p{0.15\linewidth}|}
			\hline 
			Position & Wight factor  &$ \chi $$ ^2 $ & R$ _{p} $& R$ _{wp} $  
			\tabularnewline
			\hline 
			AOT and BOT  &50 $ \% $  &3.96 &5.33&7.60
			\tabularnewline
			\hline	
			AOH and BOH &50 $ \% $&3.85 &5.23 &7.49 
			\tabularnewline
			\hline
		\end{tabular}
	\end{center}
	\caption{ Comparison of different agreement parameters obtained from the Reitveld refinement for two different combinations of Pt substitution in Mn$ _{2.7} $Pt$ _{0.3}$In.} 
	\label{Configurations4}
\end{table}

\begin{figure} [t]
	\begin{center}	
		\includegraphics[angle=0,width=8 cm, clip=true]{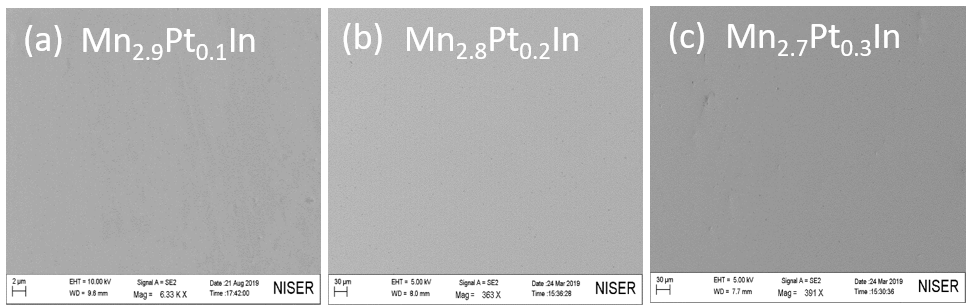}
		\caption{\label{FIG9}(Color online)SEM images of Mn$ _{3-x} $Pt$ _{x} $In}
	\end{center}
\end{figure}

\subsection{SEM and EDAX Results}
To check the homogeneity and phase purity we performed the SEM study on all the samples. Fig. 9(a)-9(c) represents the SEM images of Mn$ _{3-x} $Pt$ _{x} $In and Fig. 10(a) and 10(b) for  Mn$ _{3-y} $Ni$ _{y} $In. Usually, any spatial chemical inhomogeneity appears as spatial color (dark and bright) contrast. The EDAX analysis gives the compositional value around the exact stoichiometric requirement which can be seen from Table-\ref{Configurations5}.
Further, we have performed elemental mapping of Mn$ _{2.8} $Pt$ _{0.2} $In and Mn$ _{2.8} $Ni$ _{0.2} $In samples as shown in Fig. 11(a)-11(f) and Fig. 12(a)-12(f) for Mn$ _{2.8} $Pt$ _{0.2} $In and Mn$ _{2.8} $Ni$ _{0.2} $In respectively. The SEM images show the single-phase nature of these samples. The EDS analysis gives the atomic percentage of the constituent elements which are very close to the initially selected composition of the respective samples [see the right inset of Fig. 11(b) and Fig. 12(b)].  The individual elemental mapping was obtained from the selected area marked by the red rectangle in Fig. 11(a) and Fig. 12(a). The elemental mapping in Fig. 11(c)-11(f) and Fig. 12(c)-12(f) describes the homogeneity of the samples. All these analyses along with XRD investigation resembles the homogeneity and single-phase nature of the compounds under study.
\begin{figure} [t]
	\begin{center}	
		\includegraphics[angle=0,width=8 cm, clip=true]{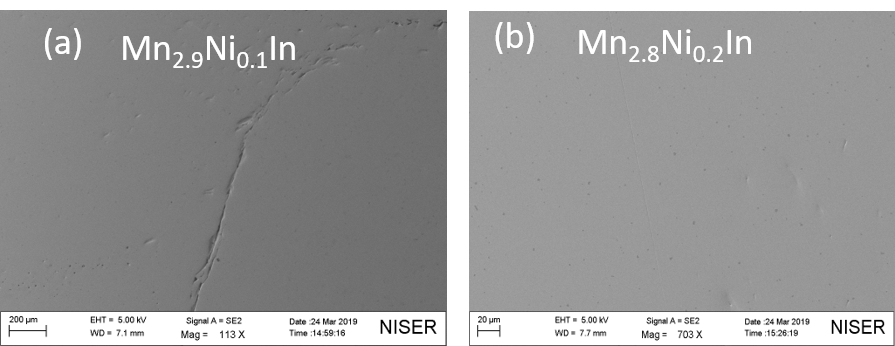}
		\caption{\label{FIG10}(Color online) SEM images of Mn$ _{3-y} $Ni$ _{y} $In}
	\end{center}
\end{figure}
\section{Magnetic Properties Studies}
\textit{M}(T) measurement for both the series of samples  show bifurcation between the the FC and ZFC curve at low-temperature regions as a result of FM cluster formation during the field cooling process. Such bifurcation sustains even up to a high field of 5 T as can be seen from the $ M $(T) data at different values of external magnetic field for Mn$ _{2.8} $Pt$ _{0.2} $In and Mn$ _{2.8} $Ni$ _{0.2} $In samples [see Fig. 13 and Fig. 14]. Also, we have carried out the ac susceptibility measurement for Mn$ _{2.8} $Pt$ _{0.2} $In and Mn$ _{2.8} $Ni$ _{0.2} $In samples. The variation of the real part of ac susceptibility ($ \chi $$ ^{'} $) with respect to temperature at various drive frequencies are shown in Fig. 15(a) and 15(b). The $ \chi $$ ^{'} $(T) shows a peak around the ordering temperature T$ _{N} $.
\begin{figure} [t]
	\begin{center}	
		\includegraphics[angle=0,width=8 cm, clip=true]{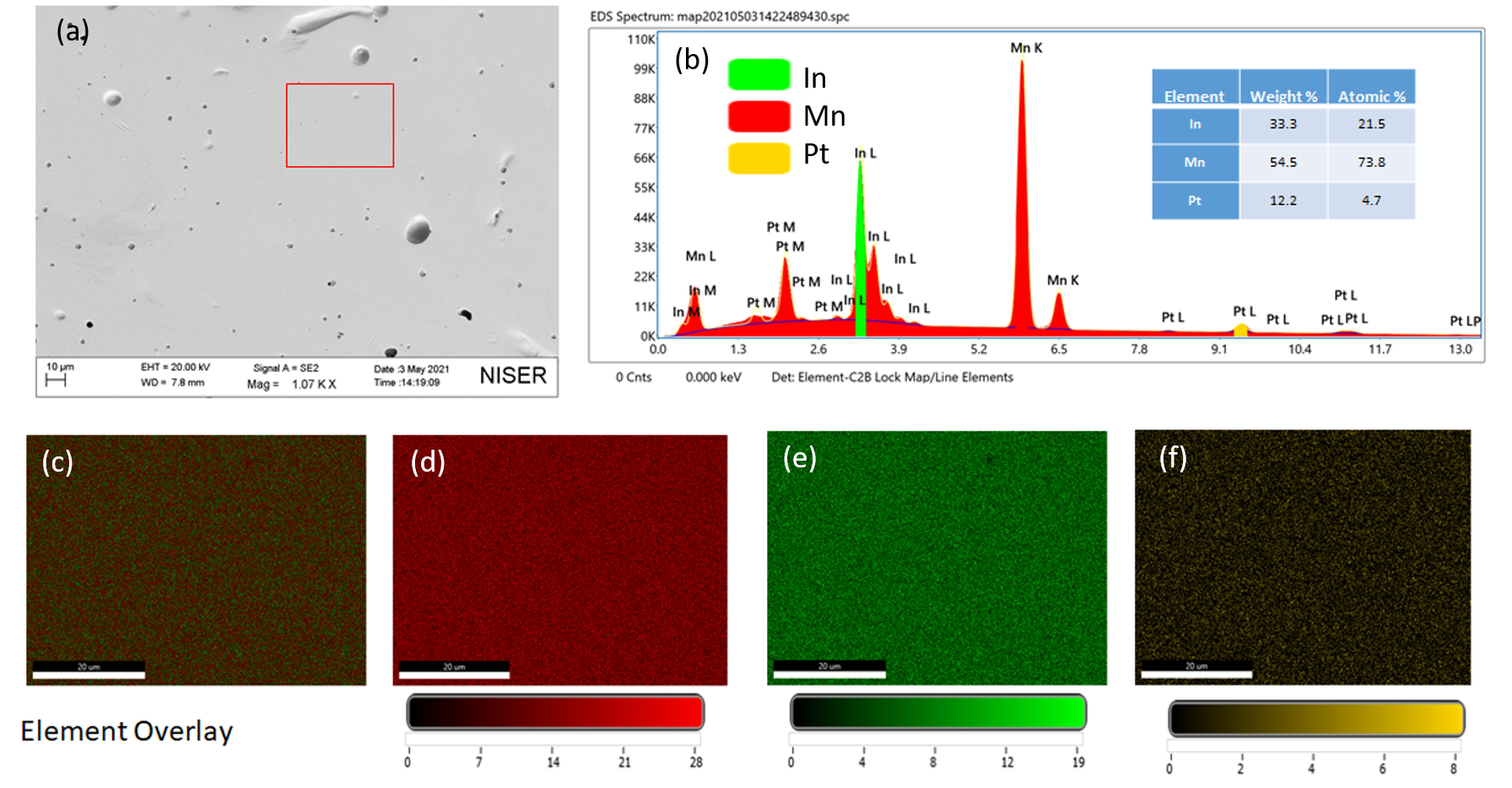}
		\caption{\label{FIG11}(Color online) Elemental mapping of the Mn$ _{2.8} $Pt$ _{0.2} $In sample. (a) SEM image of the Mn$ _{2.8} $Pt$ _{0.2} $In sample. (b) EDS spectrum of the sample. The left inset of (b) represents the color code of the individual element. The right inset of (b) represents the atomic percentage of the constituent element in this compound. (c) Elemental overlay. (d)- (f) Individual elemental mapping of Mn, In, Pt respectively. The EDS spectrum and the elemental mapping have been taken from the selected area marked in the red rectangle in (a).}
	\end{center}
\end{figure}

\begin{table}[t]
	\begin{center}
		
		\begin{tabular}{|p{0.25\linewidth}|p{0.35\linewidth}|p{0.35\linewidth}|}
			\hline 
			Comosition & Require value ( $ \% $)&Obtained from EDAX ($ \% $) 
			\tabularnewline
			\hline 
			Mn$ _{2.9} $Pt$ _{0.1} $In  &Mn-72.5, Pt-2.5, In-25   &Mn-71.6, Pt-2.6, In-25.7 
			\tabularnewline
			\hline	
			Mn$ _{2.8} $Pt$ _{0.2} $In &Mn-70.0, Pt-5.0, In-25 &Mn-67.9, Pt-5.2, In-26.9 
			\tabularnewline
			\hline
			Mn$ _{2.7} $Pt$ _{0.3} $In &Mn-67.5, Pt-7.5, In-25 &Mn-66.1, Pt-7.5, In-26.33 
			\tabularnewline
			\hline
			Mn$ _{2.9} $Ni$ _{0.1} $In &Mn-72.5, Ni-2.5, In-25&Mn-70.8, Ni-2.6, In-25.0
			\tabularnewline
			\hline
			Mn$ _{2.8} $Ni$ _{0.2} $In &Mn-70.0, Ni-5.0, In-25 &Mn-68.9, Ni-5.0, In-25.1 
			\tabularnewline
			\hline
		\end{tabular}
	\end{center}
	\caption{ Comparison of average elemental percentages obtained from EDAX with the exact stoichiometric requirement.} 
	\label{Configurations5}
\end{table}

\begin{figure} [t]
	\begin{center}	
		\includegraphics[angle=0,width=8 cm, clip=true]{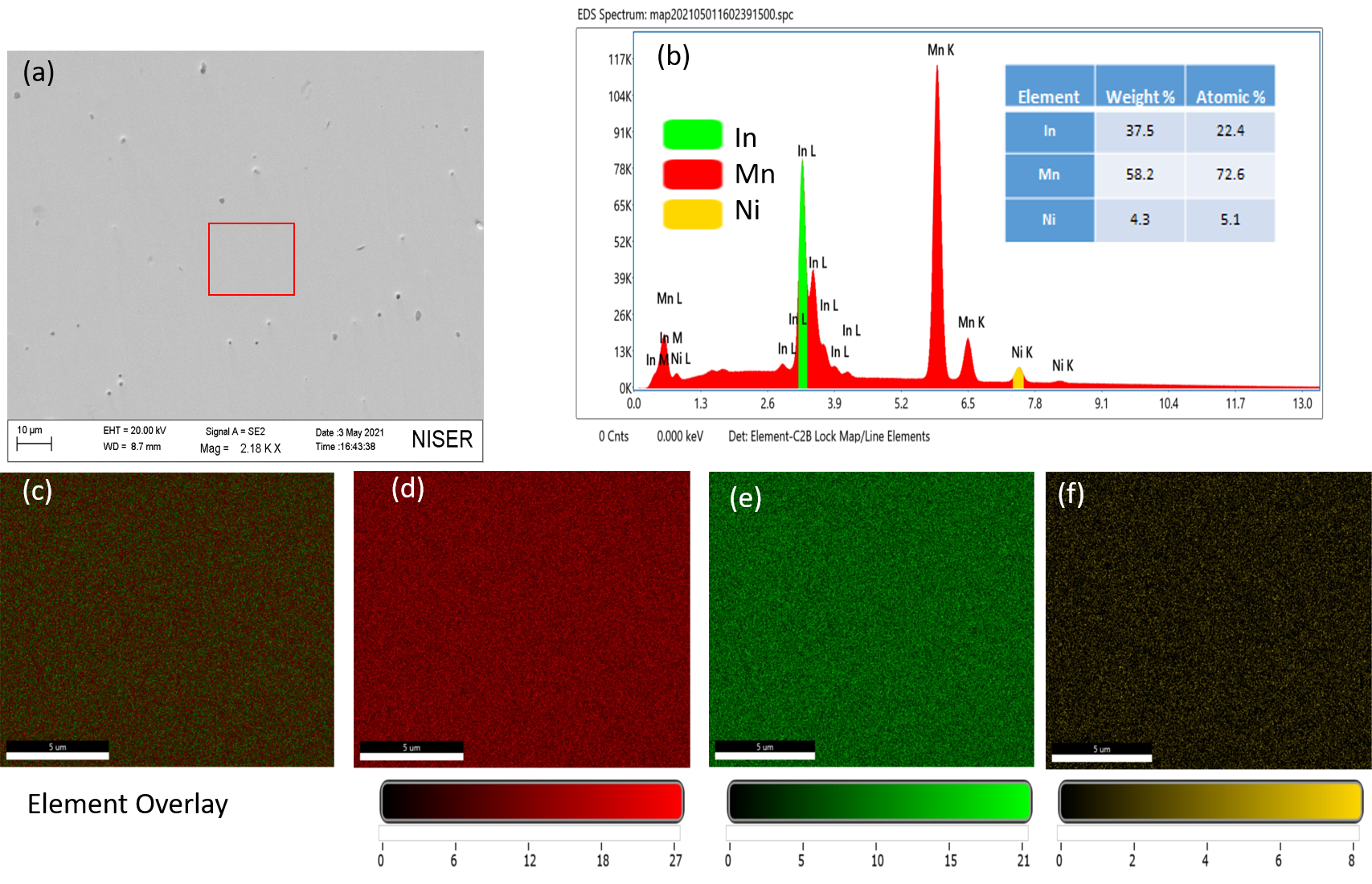}
		\caption{\label{FIG12}(Color online) Elemental mapping of the Mn$ _{2.8} $Ni$ _{0.2} $In sample. (a) SEM image of the Mn$ _{2.8} $Ni$ _{0.2} $In sample. (b) EDS spectrum of the sample. The left inset of Fig (b) represents the color code of the individual element. The right inset of (b) represents the atomic percentage of the constituent element in this compound. (c) Elemental overlay. (d)- (f) Individual elemental mapping of Mn, In, Ni respectively. The EDS spectrum and the elemental mapping have been taken from the selected area marked in the red rectangle in (a).}
	\end{center}
\end{figure}
\section{Transport Properties}
We have carried out the electrical transport measurement with a rectangular shape Hall bar. Fig. 16(a) and 16(b) represents the temperature dependence of longitudinal resistivity ($ \rho $) of  Mn$ _{3-x} $Pt$ _{x} $In and Mn$ _{3-y} $Ni$ _{y} $In in the temperature range of 2.5 -310 K. The resistivity of  Mn$ _{3-x} $Pt$ _{x} $In and Mn$ _{3-y} $Ni$ _{y} $In  samples decreases with increasing temperature i.e. a negative temperature coefficient of the resistivity. Whereas, pure Mn$ _{3} $In sample showed similar behavior only above the spin-glass transition temperature and below that found to be a metallic behavior \cite{Zhang2010,Chatterjee2020}. We have found the metallic behavior for Mn$ _{2.7} $Pt$ _{0.3}$In sample only. Fig. 16(c) shows the magnetoresistance (MR) effect of Mn$ _{2.8} $Ni$ _{0.2} $In sample at 5 K temperature.  The percentage of MR is defined as 100 x ($ \rho $$ _{H} $-$ \rho $$ _{0} $)/($ \rho $$ _{0} $), where $ \rho $$ _{H} $ is the resistivity at magnetic field H and $ \rho $$ _{0} $ is the resistivity at zero magnetic field. 
\begin{figure} [t]
	\begin{center}	
		\includegraphics[angle=0,width=8 cm, clip=true]{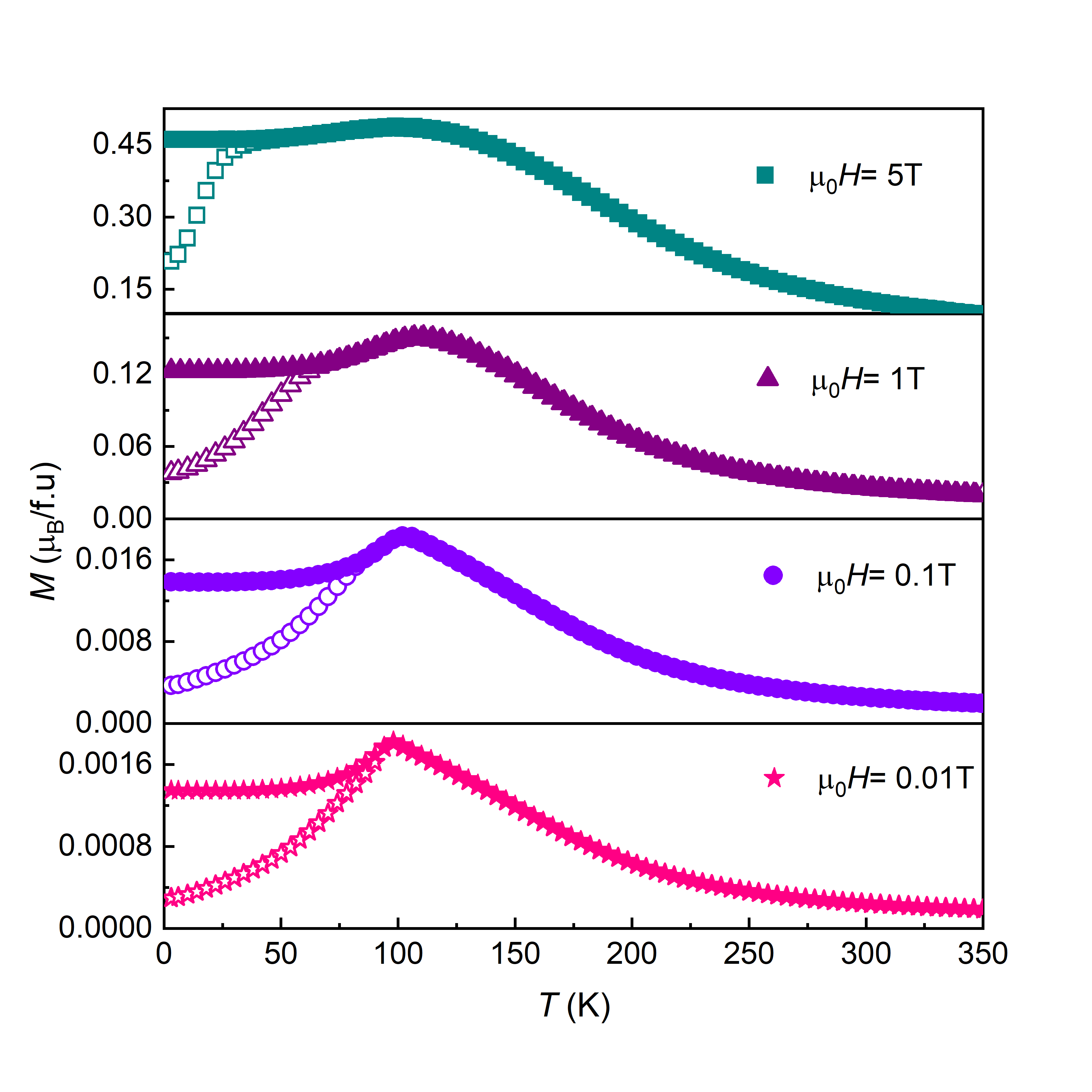}
		\caption{\label{FIG13}(Color online) The temperature-dependent magnetization plots of Mn$ _{2.8} $Pt$ _{0.2} $In sample at various applied magnetic fields. The open and solid symbols represent the zero-field cooled (ZFC) and field cooled (FC) data respectively.  }
	\end{center}
\end{figure}

\begin{figure} [t]
	\begin{center}	
		\includegraphics[angle=0,width=8 cm, clip=true]{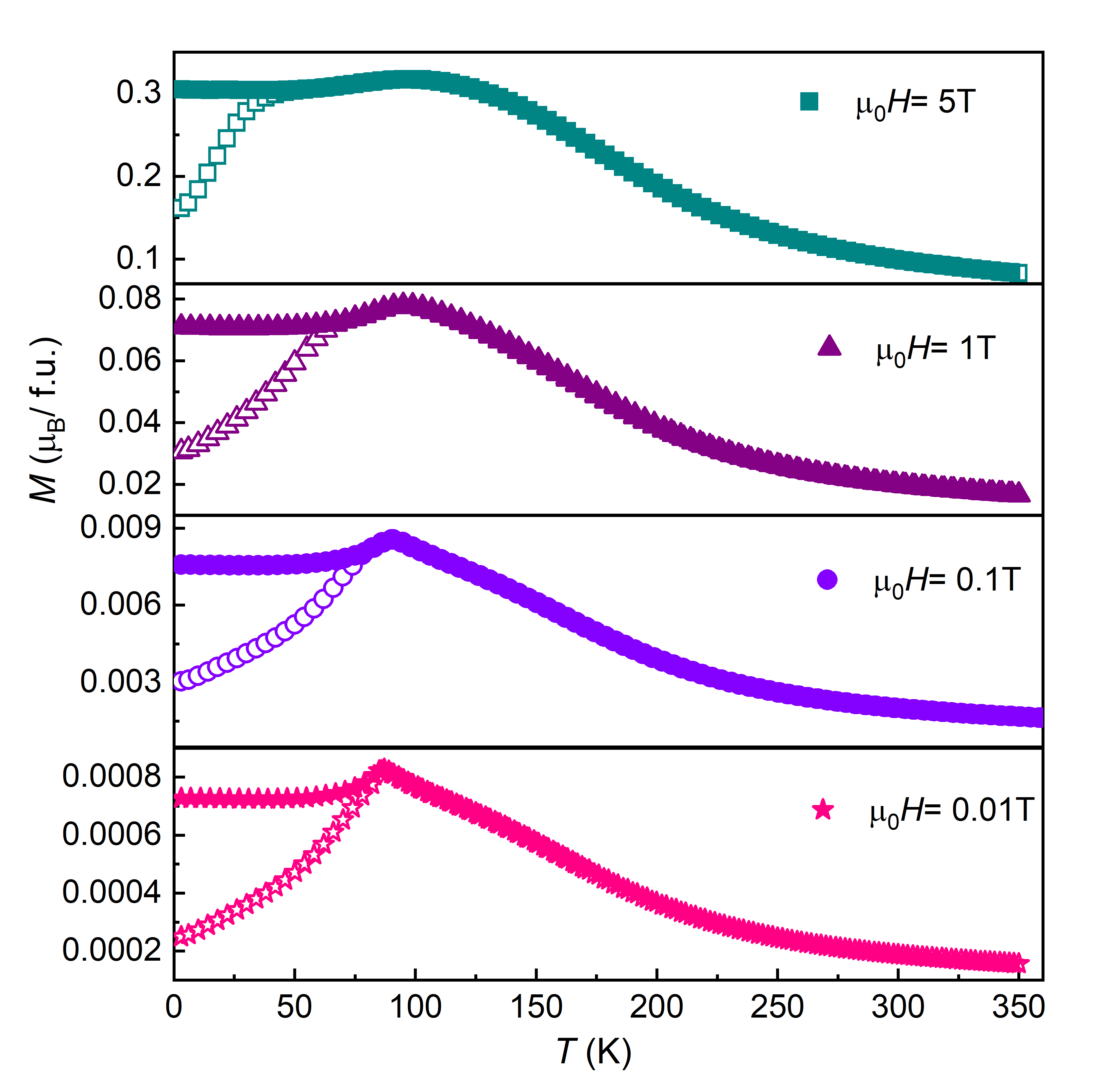}
		\caption{\label{FIG14}(Color online) The temperature-dependent magnetization plots of Mn$ _{2.8} $Ni$ _{0.2} $In sample at the various applied external magnetic field. The open and solid symbols represent the zero-field cooled (ZFC) and field cooled (FC) data respectively.}
	\end{center}
\end{figure}
\begin{figure} [t]
	\begin{center}	
		\includegraphics[angle=0,width=8 cm, clip=true]{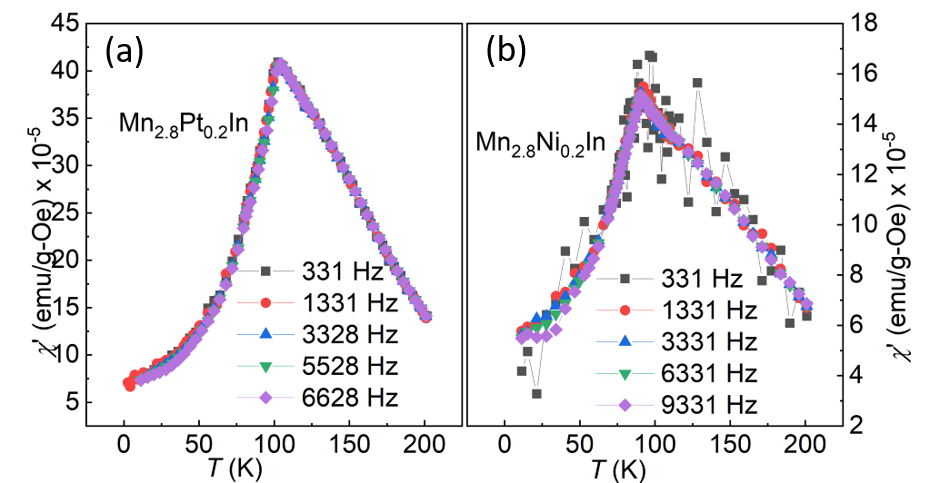}
		\caption{\label{FIG14}(Color online) Temperature-dependent of real part of ac-susceptibility ($\chi $$ ^{'} $(T)) of  Mn$ _{2.8} $Pt$ _{0.2} $In (a) and  Mn$ _{2.8} $Ni$ _{0.2} $In (b). An ac drive field of 10 Oe amplitude was used.}
	\end{center}
\end{figure}
\section{Discussion about Extra Hall Effect (EH)}
In our study, the Hall Effect experiments were performed with rectangular shape Hall bars. Also, we have performed the M(H) loop with the same sample used for Hall measurement following the same protocol and condition applied during Hall resistivity measurement. The experimental Hall resistivity ( $ \rho $$ _{yx} $) and the M-H plot are shown in Fig. 17(a) and 17(b) for $ \pm $7 T field cooled case . Also, we have performed the zero field cooled Hall resistivity for the sample Mn$ _{2.8} $Ni$ _{0.2} $In [Fig. 17(c)]. The total Hall resistivity generally consists of two terms in trivial ferromagnetic/ferri-magnetic samples and can be expressed as $ \rho $$ _{yx} $= $ \rho $$ _{N} $ + $ \rho $$ _{AH} $, where $ \rho $$ _{N} $  and $ \rho $$ _{AH} $ are the normal and anomalous Hall (AH) resistivity respectively. Normal Hall ($ \rho $$ _{N} $ )  is given by $ \rho $$ _{N} $ =R$ _{0} $H, where R$_{ 0} $ is the normal Hall coefficient and related to the carriers density of the materials and H is the external magnetic field. AH resistivity arise from the intrinsic mechanism varies as the square of the longitudinal resistivity ($ \rho $$ _{xx} $) and  directly scales with the magnetization (M) of the sample and can be written as $ \rho $$ _{AH} $=b$ \rho $$ ^2 $$ _{xx} $M, where b is the constant. So, $ \rho $$ _{yx} $ can be written as  $ \rho $$ _{yx} $= R$ _{0} $H + b$ \rho $$ ^2 $$ _{xx} $M at the high field and the unknown constant R$ _{0} $ and b is obtained by a straight line (y=mx+c) fitting in the plot of $\rho $$ _{yx} $/H versus ($ \rho $$ ^2 $$ _{xx} $M)/H. The obtained parameters (R$_{ 0} $ and b) are utilized to calculate the complete Hall loop by using the formula $ \rho $$ _{yx} $= R$ _{0} $H + b$ \rho $$ ^2 $$ _{xx} $M, where M is experimentally measured quantity i.e M(H). We have projected the calculated Hall resistivity data on the experimentaly obtain Hall data. We do not see any additional kind of Hall contribution in the ZFC Hall measurements [see Fig. 17(c)]. As can be seen form the Fig. 7 of the main manuscript that the calculated Hall resistivity matches perfectly with the experimental data at the high fields and a clear deviation appears at the first magnetization reversal region only. The calculated Hall resistivity (taking the experimentally measured M(H) into account) was subtracted from the experimental Hall data to obtain any extra Hall (EH) contribution.

\begin{figure} [t]
	\begin{center}	
		\includegraphics[angle=0,width=8 cm, clip=true]{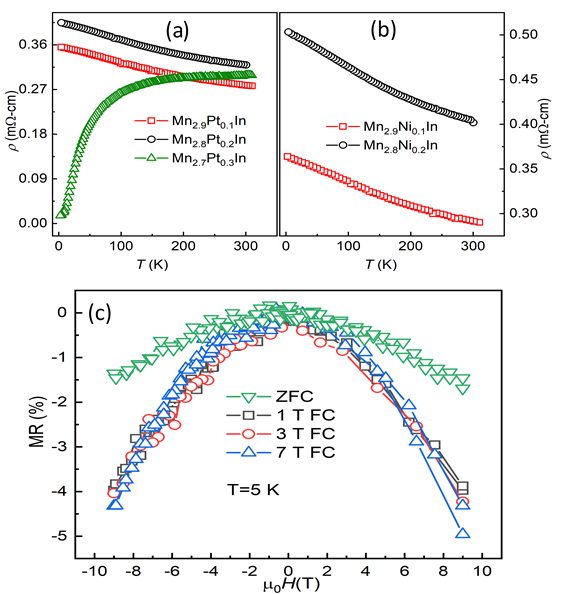}
		\caption{\label{FIG15}(Color online)Temperature-dependent of longitudinal resistivity ($ \rho $) of Mn$ _{3-x} $Pt$ _{x} $In (a) Mn$ _{3-y} $Ni$ _{y} $In (b).  (c) MR at different field cooled (FC) conditions for Mn$ _{2.8} $Ni$ _{0.2} $In.}
	\end{center}
\end{figure}
\begin{figure} [t]
	\begin{center}	
		\includegraphics[angle=0,width=8 cm, clip=true]{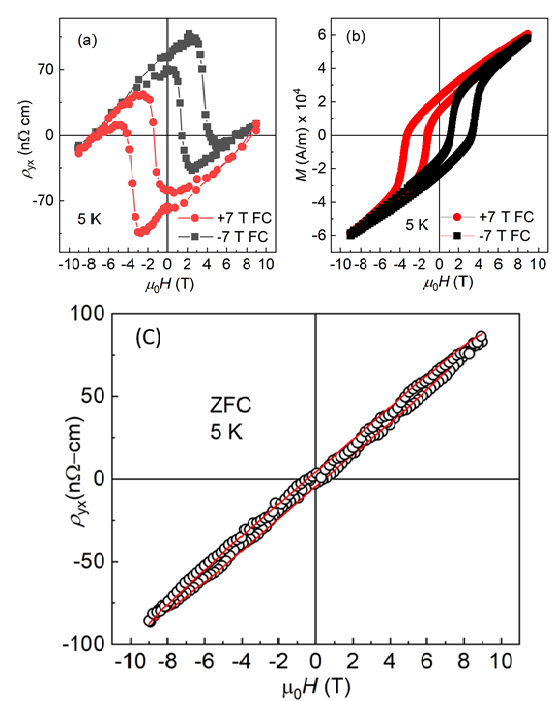}
		\caption{\label{FIG15}(Color online) Data of Hall resistivity ($ \rho $$ _{yx} $) (a) and M(H) (b) of Mn$ _{2.8} $Ni$ _{0.2} $In.  (c) Zero field cooled (ZFC) Hall resistivity data (black open circles) of Mn$ _{2.8} $Ni$ _{0.2} $In sample measured at 5 K. The red solid line is calculated Hall resistivity data. The calculated data projected onto the experimental data to see any extra Hall contribution.}
	\end{center}
\end{figure}
\begin{figure} [t]
	\begin{center}	
		\includegraphics[angle=0,width=8 cm, clip=true]{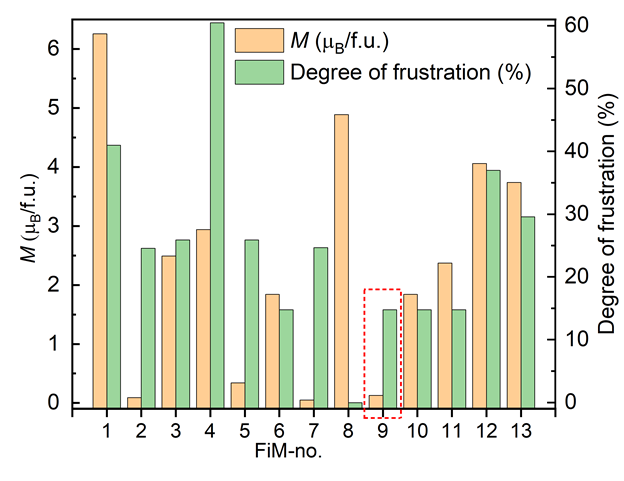}
		\caption{\label{FIG15}(Color online) The degree of frustration and the net cell moment for the 13 FiM configurations.}
	\end{center}
\end{figure}
The extra Hall (EH) may arise from the different mechanisms such as disorder, multichannel transport, or any chiral magnetic structure e.g. skyrmion/anti-skyrmion. It is well known that the spin of the conduction electron passing through a magnetic material tries to align with the local magnetization vector due to the large Hund's coupling. Hence, an electron moving through a non-collinear spin structure with non-zero scalar spin chirality experiences a large fictitious magnetic field, giving rise to an extra component to the Hall Effect, called topological Hall Effect (THE). The basis of THE can be associated with a non-vanishing scalar spin chirality $ \chi $$ _{ijk} $ = S$ _{i} $.(S$ _{j} $xS$ _{k} $), that corresponds to the solid angle subtended by three spins S$ _{i} $, S$ _{j} $, andS$ _{k} $ on a unit sphere. Therefore, THE can arise from a conventional non-coplanar magnetic ordering without the presence of any topological objects \cite{Surger2014, Wang2019,Giri2020,Rout19}.
Usually, the bump/dip kind of signature in Hall resistivity, which cannot be account for by the normal and anomalous Hall contribution is often signified as THE. However, without further evidence of topological object in the system under investigation diminishes the originality of the claim of THE. In this case, the observed bump/dip kind of signature in Hall resistivity can be readdressed in the shades of multichannel transport. More rigorously speaking, overlapping of AHE with opposite sign (one with +ve and another with –ve) can produce a signal similar to THE \cite{Gerber2018,Das2018,Kim2020,Kimbell2020,Fu2020,Wu2020}. In such cases, the opposite polarity of AHE with different coercive fields (H$ _{c} $) and/or magnetization (M) occurs as a result of magnetic inhomogeneity. Also, the sign reversal in the AHE can be induced by varying the strength of the disorder \cite{Das2018,Fu2020} in the system. This kind of magnetic inhomogeneity sometimes appears in the magnetic measurement like a step or dip kind of feature in M(H), broadening of longitudinal resistivity below the ordering temperature,  signature of hysteresis, and step-like signal in the MR, etc. However, in our system, we don’t see any such ambiguity. Also, the XRD, SEM, and EDS study establishes the structurally single-phase nature and chemical homogeneity of the sample under investigation. Therefore, the observed EH in the present system most likely arises from the non-coplanar structure at the interface.
We invoke some of the argument to validate the claim of the non-coplanar interface spin for observing THE.  In the present system, the zero fields cooled (ZFC) M-H is linear with field and the value of magnetization is very small, suggesting a magnetic compensation. This is further supported by our DFT calculation, where we establish a completely compensated collinear ferrimagnetic state. However, the field cooled (FC) M-H data show a hard ferromagnetic signature with sharp transition at the field reversal region. This behavior can be related to the magnetization rotation of uncompensated interface spins which are the backbone of inducing the exchange bias effect. In this scenario, these uncompensated interface spin becomes maximal frustrated at the coercive field region. And we find an extra Hall contribution at the region of concern after the standard method of extraction of any extra Hall contribution. Therefore, we tried to provide a possible spin configuration of the interface based on the experimental result of EH. Where we have shown a possible non-coplanar spin arrangement (as they are the basis of THE) within a single domain AFM and multi-domain state. In such a case a non-coplanar structure subtending a finite solid angle may result in a THE.
\section{DFT calculations}
Here the degree of frustration indicates the deviation from the antiferromagnetic coupling between the two nearest neighboring Mn spins. The percentage of the total number of nearest neighbors which are frustrated divided by the total number of nearest neighbors for all Wyckoff sites gives the $ \%  $ degree of frustration for a given configuration. It is clear from Fig. 18., that there is no direct correlation between the degree of frustration and the net magnetization of the unit cell. This may be due to the dominating effect of frustration arising from higher neighboring sites.

\end{document}